\documentclass[twocolumn]{aastex63}

\usepackage{amsmath,amssymb}
\usepackage[utf8]{inputenc}
\usepackage[english]{babel}
\usepackage{xcolor}
\newcommand{\unit}[1]{{\rm \, #1}}
\renewcommand{\b}[1]{\boldsymbol{#1}}
\newcommand{\cs}{c_{\rm s}}  
\newcommand{\kB}{k_{\textsc{b}}}  

\begin{document}

\title{Tidally distorted barytropes and their Roche limits, with application to WASP-12b}
\correspondingauthor{Victoria Antonetti}
\email{va4@princeton.edu}
\author{Victoria Antonetti}
\affiliation{Princeton University Observatory\\ Ivy Lane\\ Princeton NJ 08540}
\author[0000-0002-6710-7748]{Jeremy Goodman}
\affiliation{Princeton University Observatory\\ Ivy Lane\\ Princeton NJ 08540}
\shorttitle{\bf Roche limits of polytropes}
\shortauthors{Antonetti \& Goodman}

\begin{abstract}
The hot Jupiter WASP-12b has been found to be on a decaying 1.09-day orbit. The mean density of the planet inferred from transit and radial-velocity data is near its Roche limit; just how near depends on the planet's uncertain internal structure.  There is also spectroscopic evidence of mass loss.  We accurately calculate the Roche density on the assumption of a synchronously rotating $n=1$ polytrope, and find this to be only 15-20\% below the observational estimates for the mean density.  We discuss the implied shape of the planet, its lifetime before complete disruption, and its current rate of mass loss based on our improved tidal model. The present mass-loss rate is at least as sensitive to the opacity and temperature profiles of the planet's atmosphere as to its internal structure, however.
\end{abstract}

\section{Introduction}
This work is motivated by the hot Jupiter WASP-12b, which uniquely among hot Jupiters has been found to be on a decaying orbit \citep{Yee+Winn+2020}, and which shows spectroscopic indications of possible mass loss \citep{Fossati+2010}. As discussed below, the photometric and radial-velocity data suggest that the planet is close to contact with its Roche lobe, but this depends upon the planet's radial mass profile: two limits in which Roche contact can be estimated analytically give contrasting answers.

The mean density of the host star can be estimated directly from the planet's orbital period ($P$) and transit light curve \citep{Seager+2003}.  The light curve also provides the ratio $R_p/R_*$ of planetary and stellar radii.  Given the mass ratio $q\equiv M_p/M_*$ from radial-velocity data, one also has the mean density of the planet, $\bar\rho_p=3M_p/4\pi R_p^3$. For a given planetary structure, there is a critical value of the dimensionless group $G\bar\rho_p P^2$ below which Roche overflow occurs, presuming that the planet's orbit is circular and its rotation tidally locked.  \citet{Collins+2017} find $\bar\rho_p = 0.266\pm0.015\unit{g\,cm^{-3}}$ for WASP-12b.  This is on the low
side of earlier values from the literature that they tabulate in their Table~4.  A weighted average of all of these estimates based on their quoted errors is
$\bar\rho_p = 0.283\pm0.012\unit{g\,cm^{-3}}$. We emphasize that throughout this paper, $\bar\rho_p$ is defined in terms of the transit radius.  Because of the planet's tidal distortion, its volume must be somewhat larger than $4\pi R_p^3/3$, and its true mean density averaged over that volume somewhat less than $\bar\rho_p$.

If one estimates the Roche lobe in the usual way as if all of the mass of the planet were concentrated at its center, then the projected area of the lobe at mid-transit would be approximately $1.296 R_{\rm H}^2$, where $R_{\rm H}\equiv (q/3)^{1/3}a$ is the Hill radius and $a$ the orbital semimajor axis.  This would correspond to a critical transit radius
$R_{p,\rm crit}\approx 0.642 R_{\rm H}$.  Given the observed orbital period $P=1.09142\unit{d}$ \citep{Yee+Winn+2020}, this model for the planet's structure
predicts a minimum mean density $\bar\rho_p\approx 2.71 G^{-1}(2\pi/P)^2\approx 0.180\unit{g\,cm^{-3}}$: significantly below the observed value.  Thus on this model, the planet would be safely within its Roche lobe.

If instead the planet had uniform density, then the limiting equipotential at Roche contact would be ellipsoidal with semiaxes $b_1:b_2:b_3=1 : 0.511 : 0.483$, and $b_1b_2b_3= 0.06757 qa^3$ \citep[e.g.][]{Chandrasekhar1987}.  The corresponding mean density based on the transit radius $\sqrt{b_2 b_3}$ would then be $0.473\unit{g\,cm^{-3}}$: well \emph{above} what is inferred from the data.
On this model, the planet would already have been tidally disrupted.
The minimum density $\bar\rho_{p,\rm crit}$ for a realistic planetary structure lies somewhere between these limiting cases.  

To a first approximation, the radial density profile of a jovian planet can be described as an $n=1$ Emden polytrope \citep{Stevenson2020}.  One would have thought that the tidal limit of such polytropes would have been worked out long ago. \cite{Chandrasekhar1933} has calculated weak tidal distortions for such bodies by perturbative methods, while \cite{Rasio+Shapiro1992,Rasio+Shapiro1994} have obtained numerical results via smooth-particle hydrodynamics for mass ratios $q\equiv M_2/M_1$ near unity, as they were motivated by neutron-star mergers.  \cite{Taniguchi+2002} have made calculations for $n=1$ (adiabatic index $\gamma=2$) and mass ratios as small as $q=0.1$.  Numerical difficulties prevented them from making models in full Roche contact, but by double extrapolation of their tabulated results in both $q$ and tidal strength, we estimate axis ratios
$b_1:b_2:b_3\approx 1:0.62:0.59$ for such a case when $q=10^{-3}$.  At WASP-12b's orbital period, this would imply $\bar\rho_{p,\rm crit}\approx 0.24\unit{g\,cm^{-3}}$, 10-15\% below the observational estimates quoted above; in other words, the planet would not yet fill its Roche lobe.  In view of this slender margin and the importance of the WASP-12 system, a more careful calculation would seem to be called for.  That is the goal of the present paper.

Section 2 describes our implementation of the Ostriker-Mark (OM) method for  tidally distorted polytropes and some tests against analytic results.
Section 3 and Table~\ref{tab:roche-table} present  Roche-limited models for Emden indices $n\in\{1,3/2,3\}$ models and mass ratios $q=M_{\rm p}/M_*\in\{10^{-4},10^{-3},10^{-2}\}$.
Section 4 and the Appendix discuss mass-loss rates for models that best fit the observed $\bar\rho_{\rm p}$; these do not quite fill the Roche lobe but we endow them with realistic finite-temperature atmospheres, which we borrow from recent work in spherical symmetry. We also estimate from the observed orbital-decay rate \citep{Yee+Winn+2020} and our best-fitting polytropic models how long it will be before the main body of the planet (i.e., its roughly isentropic core) comes into Roche contact. Section~5 sums up and points to directions for future work.

\section{Methodology}

Our general problem is determining the shape of a small fluid body (jovian planet) in circular orbit around a larger body (the star).  WASP-12b's orbit is indeed consistent with circular \citep{Yee+Winn+2020}.  Presumably this is enforced by tidal dissipation within the planet, in which case the planet's (unobserved) spin is expected to be synchronous apart from thermally or convectively driven zonal flows. Even with these simplifications to the orbital dynamics, a very accurate calculation of the equilibrium shape would require accurate knowledge not only of the equation of state, but also of the planet's internal stratification of entropy and composition. 

These things being uncertain even for Jupiter, let alone WASP-12b, we represent the planet as a polytrope, i.e. pressure and density are locally related by $p(\rho)=K\rho^{(n+1)/n}$ for some constant $K$. For the most part, we take $n=1$. This is believed to be a rather good first approximation to Jupiter's equation of state \citep{Stevenson2020}. Furthermore, the Juno mission has measured Jupiter's quadrupolar Love number $k_{22}=0.565\pm0.006$ \citep{Durante+2020}; for a hypothetical non-rotating planet with Jupiter's radial density profile, this would be reduced by about 9\%, i.e. to $k_2\approx 0.51$.  An $n=1$ polytrope has $k_2=15/\pi^2\approx 0.520$, whereas $k_2=3/2$ for $n=0$, $k_2\approx 0.285$ for $n=3/2$ and of course $k_2\approx 0$ for a strongly centrally concentrated body. While Jupiter's internal structure is clearly more complicated than a polytrope, Love numbers are a direct measure of tidal response, albeit for infinitesimal tidal fields, and depend only on the radial density profile.  On the other hand, WASP-12b is substantially inflated, with a mean density less than a quarter of Jupiter's, so that its internal structure ($d\ln P/d\ln\rho$) may be rather different from Jupiter's.  Therefore we have considered other values of the polytropic index, especially when estimating the planet's mass-loss rate (Appendix~\ref{sec:Mdot}).

We use the Ostriker-Mark method for finding the self-consistent density and potential for the planet in the presence of the tidal field exerted by the star \citep{Ostriker+Mark1968}.  Some minor additions to the method are needed to cope with spherical harmonics of odd degree, which were not needed in the original application to rotating stars.  For completeness, we briefly review the method here, with emphasis on our additions to it.

Since we assume the planet to rotate synchronously at angular velocity $\b{\Omega}$, hydrostatic equilibrium in the corotating frame takes the form
\begin{equation} \label{eq:hydro}
{\nabla p }/\rho + \nabla \Psi = 0
\end{equation}
in terms of an effective potential
\begin{equation} \label{eq:psidef}
\Psi = \Phi_p + \Phi_* - \tfrac{1}{2} [\b{\Omega \times} (\b{r-r}_0)]^2\,,
\end{equation}
in which
\begin{equation} \label{eq:grav}
\Phi_p(\b{r}) = -G \int\frac{\rho_p (\b{r'})}{ |\b{r}-\b{r'}| }\mathrm{d^3}\b{r'} \ +\  \mbox{constant}
\end{equation}
is the gravitational potential of the planet, $\Phi_*$ is that of the star, represented here by a point mass $M_*$, and $\b{r}_0$ is the barycenter of the orbit.

It follows from eq.~\eqref{eq:hydro} that the pressure is constant on equipotentials $\Psi=\mbox{constant}$; furthermore from
the curl of the equation, $\b{\nabla}\rho\b{\times\nabla}p=0$, so that the density is also constant on equipotentials.
Hence $p$ and $\rho$ may be regarded as functions of $\Psi$ and therefore of one another. One may introduce a
formal enthalpy function
\begin{equation} \label{eq:Hdef}
H (p) =  \int_0^p \frac{\mathrm{d}p'}{\rho(p')}\,,
\end{equation}
in terms of which hydrostatic equilibrium can be re-expressed as
\begin{equation} \label{eq:hydro1}
H + \Psi = \Psi_s = \mbox{constant,}
\end{equation}
$\Psi_s$ being the value of $\Psi$ at the surface of the planet where its pressure vanishes.

Note that eqs.~\eqref{eq:Hdef}-\eqref{eq:hydro1} are consequences of hydrostatic equilibrium regardless of the equation of state, which in general involves more than one independent thermodynamic variable (e.g. density, entropy, and chemical composition).
To use the OM method, however, one needs an explicit formula or tabulation for $p(\rho)$.
A polytropic relation $p=K\rho^{(n+1)/n}$ leads to
\begin{align} \label{eq:poly}
\rho(\Psi) &= \left(\frac{\Psi_s-\Psi}{(n+1)K}\right)^n & \mbox{where }\Psi &< \Psi_s\nonumber\\
           &= 0 & \mbox{where } \Psi &\ge \Psi_s\,.
\end{align} 
It's convenient to choose the arbitrary constant in eq.~\eqref{eq:grav} so that $\Psi_s=0$ and so $\Psi<0$ within the planet.
For the reasons given earlier, the choice $n=1$ appears to be most relevant to jovian planets, but we also test our code against analytic results for $n=0$, and we compute models for $n=3/2$ and $n=3$ as well as $n=1$.
We put the origin of $\b{r}$ at the center of mass of the planet. 
The distance from there to the barycenter is $r_0=a/(1+q)$, with $a$ the separation between the centers of the planet and the star, and $q=M_p/M_*$ the mass ratio.
We ignore the oblateness and tidal distortion of the star (WASP-12 appears to be slowly rotating) and expand its potential
in zonal harmonics about the planet:
\begin{equation}\label{eq:Phistar}
\Phi_* = -\frac{GM_*}{a}\sum\limits_{l=1}^\infty \left(\frac{r}{a}\right)^l P_l(\cos\theta)\,.
\end{equation}
Notice that the term at $l=0$, which exerts no force on the planet, is omitted from the sum, so that $\Phi_*=0$ at $r=0$.
The polar angle $\theta$ is measured from the line between the planet and the star.  Following the conventions of the restricted three-body problem, we make this the $x$ axis, except that $x=y=z=0$ at the center of the planet rather than the barycenter, and the star lies at $(x,y,z)=(a,0,0)$; $y$ increases in the direction of the planet's orbital motion, and $z$ in the direction of $\b{\Omega}$.

With these conventions, the centrifugal part of the effective potential \eqref{eq:psidef} becomes
\begin{equation}\label{eq:centrifugal}
-\tfrac{1}{2} [\b{\Omega \times} (\b{r-r}_0)]^2 \to \Omega^2 r_0 x - \tfrac{1}{2}\Omega^2(x^2+y^2)\,,
\end{equation}
upon dropping a constant term $-\tfrac{1}{2}\Omega^2r_0^2$ that exerts no force. Together with eq.~\eqref{eq:Phistar},
this ensures that $\Psi(\b{0})=\Phi_p(\b{0})$, i.e. the effective potential at the center of the planet is due to solely to its self-gravity.
The next term, $\Omega^2 r_0 x$, is almost balanced by the $l=1$ part of the stellar potential \eqref{eq:Phistar}, but not exactly, because of the finite size and nonspherical shape of the planet: 
Let us define the $l^{\rm th}$ mass moment of the planet by
\begin{equation}\label{eq:moment}
\mu_l \equiv \int \rho(\b{r})\,r^l P_l(\cos\theta)\,\mathrm{d}^3\b{r}\,,
\end{equation}
so that $\mu_0=M_p$; also $\mu_1=0$ because we expand around the planet's center of mass.  The $(l+1)^{\rm st}$ term in the expansion \eqref{eq:Phistar} for the stellar potential $\Phi_*$ exerts (through $\partial\Psi_*/\partial x$) a net
attractive force on $\mu_l$.
Thus the balance of centrifugal and gravitational forces on the planet is
\begin{equation}\label{eq:l1balance}
    M_p\Omega^2 r_0 = \frac{GM_*M_p}{a^2} + \frac{GM_*}{a^2}\sum_{l\ge 2} \frac{(l+1)\mu_l}{a^l}\,.
\end{equation}
The left side is due to the linear part of the centrifugal potential \eqref{eq:centrifugal}; the quadratic part exerts no net force because $\mu_1=0$.  Since $r_0=a/(1+q)$, eq.~\eqref{eq:l1balance} can be rewritten as
\begin{align}\label{eq:delta}
    \Omega^2 &= \frac{G(M_*+M_p)}{a^3}\times \left[1+M_p^{-1}\sum_{l\ge 2} (l+1)\mu_l a^{-l}\right]\nonumber\\
    &\equiv \Omega_K^2\times(1+\delta)\,,
\end{align}
with $\Omega_K$ being the ``Keplerian" orbital frequency (mean motion) that would obtain if the planet were a point mass or a sphere.
The notation $\delta$ for the correction follows LRS93. For a planet of mean radius $\bar R_p\ll a$,
$\delta$ scales as $(\bar R_p/a)^2$, which becomes $O(q^{2/3})$ at Roche contact.  Since we are interested in mass ratios
$q\approx 10^{-3}$, the correction might be $\sim 1\%$; in fact it is an order of magnitude smaller
because of the central concentration of the ($n\approx 1$) planet.

Our OM iterations are carried out in Emden units: $4\pi G = (n+1)K = \rho_c=1$.  So for example, eq.~\eqref{eq:poly} becomes
simply $\rho(\Psi)=[\max(-\Psi,0)]^n$.  To emphasize that a physical
quantity is expressed in Emden units, we sometimes mark it with a tilde.
In particular, 
\begin{equation}\label{eq:tidal-parameter}
\tilde\Omega_K^2\equiv\frac{\Omega_K^2}{4\pi G\rho_c}
\end{equation}
is our dimensionless measure of the strength of the tide, corresponding to $(1+q)\tilde\mu/4$ in the notation of LRS93.

Our OM procedure is standard overall:
\begin{enumerate}
\item At the start of the $(k+1)^{\rm st}$ iteration, we have an estimate $\Psi^{(k)}$
for the effective potential.  Using this, we estimate the planet's density $\rho^{(k+1)}$ on a uniform grid in each coordinate
$(r,\theta,\phi)$ using eq.~\eqref{eq:poly} with $\Psi\to\Psi^{(k)}$.
\item At each radial grid point $r_i$, we transform the density samples on the $(\theta,\phi)$ grid to coefficients of
spherical harmonics $Y_{l,m}(\theta,\phi)$ up to some maximum degree $l_{\max}$ using the publicly available Python package
{\tt Pyshtools} \citep{shtools}.\footnote{The numbers of longitudinal and latitudinal samples demanded by {\tt Pyhshtools} scale in proportion to the maximum $l$. In order to sample the sphere more densely, we zero-pad the coefficients of the spherical harmonics up to $2l_{\max}$.}  In this way we get a discrete approximation to the series representation
\begin{equation} \label{eq11}
\rho (r,\theta,\phi) = \sum_{l=0}^{\infty} \sum_{m=-l}^{l} \rho_{l,m} (r) Y_{l,m} (\theta,\phi).
\end{equation}
\item The corresponding radial functions $\Phi_{l,m}(r)$ in the expansion of the potential $\Phi_p(r,\theta,\phi)$ are obtained
from the usual radial integrals over $\rho_{l,m}(r)$, except that the boundary condition on the $l=m=0$ component
is such that $\Phi_p(\b{0})=-1$ instead of $\Phi_p(\infty)=0.$  We approximate these radial integrals via spline interpolation
and Simpson's Rule.  We now have preliminary updates to the planet's potential $\Phi_p$ and mass moments $\mu_l$.
\item In general the center of mass of the planet is no longer exactly at the origin.
We alter the radial functions $\Phi_{l,0}(0)$ in the expansion of $\Phi_p$ slightly to correct this, as described below.
\item Using the new estimate $\mu_0^{(k+1)}$ for the planet's mass $M_p$, we re-estimate the separation of the planet
and star by solving $\tilde\Omega_K^2=M_p(1+1/q)/4\pi a^3$ for $a$ (recall $G\to1/4\pi$ in Emden units).  Then we update
$\Omega^2$ and $\delta$ using eq.~\eqref{eq:delta}.
\item We add terms to the coefficients to represent the stellar potential \eqref{eq:Phistar} and the centrifugal potential
\eqref{eq:centrifugal}.  The latter contributes at $l\in\{0,1,2\}$; the contribution to the radial function for $l=0$
is quadratic in $r$ rather than constant, because $y^2+z^2$ averages to $(2/3)r^2$ over a sphere.  With these tidal and centrifugal
additions, we have a preliminary representation for the next iterate $\Psi^{(k+1)}$ of the effective potential in the form of a
series of spherical harmonics with coefficients depending on radius.
\end{enumerate}

We elaborate here on step 4. For $q\ll 1$, the tidal-plus-centrifugal potential $\Delta\Psi\equiv\Psi-\Phi_p$ is
predominantly quadrupolar ($l=2$), apart from the centrifugal contribution at $l=0$ noted above.  If we ignore all
odd values of $l$ in $\Delta\Psi$, then step 4 in not needed.  However, the odd-$l$ components of $\Delta\Psi$
contribute to the $l=1$ parts of the planet's density distribution.  This is true even for $n=1$ where the
relation between $\rho$ and $\Psi$ is linear within the planet, but nevertheless is globally nonlinear
because of the truncation of the density at $\Psi=0$.
Physically, the distortion of the planet is stronger on the side facing the star than on its opposite side.
Thus, each iteration generally perturbs the center of mass.\footnote{But only along the $x$ axis, because
$\Delta\Psi$ and the density are always symmetric under the reflections $y\to-y$ an $z\to-z$.}
We cannot simply ignore the $(l,m)=(1,0)$ part of $\Phi_p$, because it should be nonzero within the planet.  
Thus at step 4, we find the
displacement $\Delta x=\mu_1/\mu_0$ of the center of mass, and correct for it via the first-order approximation
\begin{equation*}
    \Phi_p(x-\Delta x,y,z)\approx \Phi_p(x,y,z) - \Delta x \frac{\partial\Phi_p}{\partial x}.
\end{equation*}
Since $\partial/\partial x = \cos\theta\partial/\partial r +r^{-1}\partial/\partial\cos\theta$, the corrections to the radial functions are (to first order in $\Delta x$)
\begin{multline}\label{eq:shift}
\Delta\Phi_{l,m}(r) = -\Delta x\, C(l,m)\left(\frac{d}{dr}-\frac{l-1}{r}\right)\Phi_{l-1,m}\\
-\Delta x\, C(l+1,m)\left(\frac{d}{dr}+\frac{l+1}{r}\right)\Phi_{l+1,m}\,,
\end{multline}
in which the coefficients are
\begin{equation}
    C(l,m)\equiv \sqrt{\frac{l^2-m^2}{(2l+1)(2l-1)}} \quad\mbox{if $l\ge|m|$, else 0.}
\end{equation}
Without step 4, the method usually does not converge when the odd parts of the tide are included.  Hachisu's Self-Consistent Field
method \citep{Hachisu+Eriguchi1984}, while otherwise similar to Ostriker-Mark, has the advantage in this respect that the endpoints
of the bodies on the $x$ axis are directly constrained; however, Hachisu's method does not directly constrain the mass ratio (unless $q=1$), which has to be discovered through iteration.

\subsection{Tests of the method}

We have satisfied ourselves that the output parameters of our models presented in Table~\ref{tab:roche-table} [columns (3)-(10)] converge as we refine the radial grid and increase the maximum spherical-harmonic degree ($l_{\max}$).
In fact, values shown there are obtained by Aitken acceleration with respect to both $\Delta r\in\{0.005,0.0025,0.00125\}$ (in Emden units) and $l_{\max}\in\{8,12,16\}$.

We have also tested our code against analytic results.
Our code accurately reproduces the properties of spherical Emden polytropes.  In particular, for $n=0$ and $n=1$, the calculated masses and radii agree with the exact values to at least 5 places (in fact 7 places except for the mass of the $n=0$ model); this is for a radial grid spacing $\Delta r=0.0025$ with rational grid points that do not coincide with the exact radii ($\xi_{\max, n=0}=\sqrt{6}$ and $\xi_{\max,n=1}=\pi$): $\xi_{\max}$ is estimated by linear interpolation between neighboring grid points.

To test the accuracy of tidal distortions, we calculate the Love numbers
\begin{equation} \label{eq17}
k_l = \left.\frac{\Phi_{l,m}^{\rm self}(r)}{\Phi_{l,m}^{\rm external}(r)}\right|_{r=r_{\max}}
\end{equation}
where $\Phi_{l,m}^{\rm external}(r)$ is an infinitesimal tide $\propto r^l P_l(\cos\theta)$, $\Phi_{l,m}^{\rm self}(r)$ is the corresponding component of the distorted planet's potential, and $r_{\max}$ is the radius of the spherical planet.
In particular, the quadrupolar Love number $k_2=15/\pi^2-1\approx 0.5198$ for $n=1$, whereas our code yields $0.5208$ with $\Delta r=0.00125$ and a tidal
strength $\tilde\Omega_K^2=10^{-8}$.  This is an error of $0.2\%$, and it is linear in $\Delta r$: Aitken extrapolation based on the results for $\Delta r\in\{0.005,0.0025,0.00125\}$ yields the correct value to 5 places. 
The Love numbers for $n=0$ are also known analytically but are much more difficult for our code.
During the OM iterations, we evaluate the density from the potential on a fixed grid in $(r,\theta,\phi)$, making no attempt to interpolate between grid points.
For $n=0$, the density depends only on the sign of the potential (vanishing for $\Psi>0$); therefore, if we apply a very small tidal potential that should move the bounding equipotential a distance $<\Delta r$, there is no change to the density on the grid at all.
When we apply a finite tidal distortion $\tilde\Omega_K^2=10^{-3}\approx 0.044\tilde\Omega_{K,\rm Roche}^2$, we obtain the estimate
$k_2=1.547$, which is $3\%$ larger than the exact value $k_2=3/2$ for an infinitesimal tide.

To test the ability of the code to measure finite distortions, we search for the Roche limit when the Emden index $n=0$ and the mass ratio $q\to 0^+$.  In this case, the exact result is known via elliptic integrals (e.g. \citealt{Chandrasekhar1987}): $\tilde\Omega_{K,\rm Roche}^2\approx 0.02252325$.  To estimate this with our code, we set $q=10^{-18}$, $l_{\max}=12$, $\Delta r=0.00125$, and gradually increase
$\tilde\Omega_K^2$ until the code no longer converges.  The last converging model is $\tilde\Omega_K^2\approx0.02266$, or
$0.6\%$ more than the exact value.  The size of the error is disappointing, but as explained above, $n=0$ is more stressful for our code than $n\ge1$.

%

\section{Results}

We define the critical model for a positive polytropic index $n$ and mass ratio $q$ to be that for which the surface gravity vanishes at the pole facing the star: i.e., $\partial\Psi/\partial x=0$ at the point where surface $\Psi=0$ intersects the positive $x$ axis (``L1'').\footnote{This criterion fails for $n=q=0$ because $\Psi$ is then homogeneously quadratic in the coordinates within the body, so that if $\partial\Psi/\partial x$ vanishes at any $x\ne0$, then it vanishes throughout.} With this definition, the critical model is at the brink of mass transfer onto its companion even in the approximation that the density and temperature vanish at its surface, and therefore can be considered to be in contact with its Roche lobe. (In Appendix~A, following \cite{Jackson+2017} and earlier authors, we estimate the mass-loss rate for slightly weaker tides when a finite atmospheric temperature is allowed for.)  LRS93 have shown within their ellipsoidal approximation that secular or dynamical instability typically sets in before the Roche limit, but the differences in the tidal parameter \eqref{eq:tidal-parameter} are very small for $q\ll 1$.  Table~\ref{tab:roche-table} displays the calculated dimensionless parameters of the critical model for three values of $n$ and $q$ that seem likely to be relevant for gaseous exoplanets.  The parameters are estimated by Aitken extrapolation in $\Delta r$ and $\ell_{\max}$ and should be accurate to the number of significant digits shown.

\begin{deluxetable*}{llllllllll}
  \tablenum{1}
  \tablecaption {Polytropic models in Roche contact. \label{tab:roche-table}} 
  \tablewidth{0pt}
  \tablehead{
    \colhead{$q$} & \colhead{$\tilde M_{\mathrm{p}}$} & \colhead{$\tilde\Omega_{\mathrm{K}}^2$} & \colhead{$\delta$} & \colhead{$\tilde a_x$} & \colhead{$\tilde a'_x$} & \colhead{$\tilde a_y$} & \colhead{$\tilde a_z$} & \colhead{$\tilde\Psi_{\mathrm{L2}}$} & \colhead{$\bar\rho_{\mathrm{p}}$}
}
  \decimalcolnumbers
  \tablecolumns{10}
  \startdata
  \cutinhead{$n=1$}
$10^{-2}$ & 43.48 & 0.009355  & $2.608\times10^{-3}$ & 4.970 & 4.167 & 3.076 & 2.792 & 0.07337 & 0.2330 \\
$10^{-3}$ & 44.00 & 0.009854  & $6.469\times10^{-4}$ & 5.062 & 4.422 & 3.063 & 2.771 & 0.03597 & 0.2284 \\
$10^{-4}$ & 44.28 & 0.010100  & $1.494\times10^{-4}$ & 5.105 & 4.627 & 3.057 & 2.761 & 0.01729 & 0.2263 \\
  \cutinhead{$n=3/2$}
$10^{-2}$ & 35.92 & 0.005246  & $1.250\times10^{-3}$ & 5.505 & 4.621 & 3.623 & 3.318 & 0.05114 & 0.2076 \\
$10^{-3}$ & 36.11 & 0.005541  & $3.056\times10^{-4}$ & 5.575 & 4.875 & 3.615 & 3.299 & 0.02451 & 0.2002 \\
$10^{-4}$ & 36.20 & 0.005690  & $6.988\times10^{-5}$ & 5.606 & 5.080 & 3.612 & 3.290 & 0.01156 & 0.1966 \\
  \cutinhead{$n=3$}
$10^{-2}$ & 25.57 & 0.0005679 & $9.867\times10^{-5}$ & 10.098 & 8.450 & 6.957 & 6.468 & 0.01903 & 0.1890  \\
$10^{-3}$ & 25.58 & 0.0005990 & $2.368\times10^{-5}$ & 10.198 & 8.890 & 6.955 & 6.445 & 0.009045 & 0.1804 \\
$10^{-4}$ & 25.59 & 0.0006148 & $5.357\times10^{-6}$ & 10.239 & 9.256 & 6.955 & 6.434 & 0.004232 & 0.1762 \\
\enddata
\tablecomments{Col. (1), mass ratio $M_{\rm planet}/M_{\rm star}$; (2), planet mass;
  (3), tidal strength [eq.~\eqref{eq:tidal-parameter}]; (4), orbital frequency correction [eq.~\eqref{eq:delta}]; (5), radius at substellar  point;
  (6), antipodal radius; (7), radius in orbital direction; (8), radius normal to orbital plane; (9), effective potential at $L_2$; (10), mean
  density based on transit radius when scaled to orbital period of WASP 12b.  Columns (2)-(3) \& (5)-(9) in Emden units
  $4\pi G=\rho(0)=K=1$. Col.~(10) in $\mathrm{g\,cm^{-3}}$.}
\end{deluxetable*}

We also calculate models that match the mean densities estimated from observations of WASP 12b.  For $n\ge 1$, such models lie within their Roche lobes: the surface gravity at the inner
Lagrange point is positive, though small compared to that of a spherical planet of the same
mass and mean density.
For $n=1$, the dimensionless tidal parameters \eqref{eq:tidal-parameter}
are $\tilde\Omega_K^2=0.00760$ and $0.00691$ for mean densities of $0.266$ and $0.283\,\mathrm{g\,cm^{-3}}$, respectively.

\section{Discussion}
By modeling WASP-12b as a polytrope, we pretend that the density and temperature vanish on its limiting equipotential.
This is not physical; the surface of the planet must have a finite temperature, which will be rather high on the side facing the star: $3000-4000\unit{K}$.
The atmosphere therefore has no sharp edge, but attenuates roughly exponentially with scale height $H= \cs^2/g$,
$g$ being the surface gravity and $\cs=(\kB T/\mu m_p)^{1/2}$ the thermal speed.  In fact the atmosphere
can be expected to extend to the Lagrange points,\footnote{By this we mean the saddle points of the effective potential nearest the planet, although their positions are slightly different from those in the restricted three-body problem because we are not treating the planet as a point mass.} where it will pass through sonic points and escape the planet.  The mass-loss rate from WASP-12b has been estimated on this basis by several groups, notably \cite{Li+2010}, \cite{Lai+2010}, \cite{Bisikalo+2013}, \cite{Jackson+2017}, and \cite{Dwivedi+2019}. Our methods are detailed in Appendix~\ref{sec:Mdot}.  We use the formula of \cite{Jackson+2017} but improve upon it by replacing their assumption of a constant thermal speed with one based on recent models for the day-side atmospheric temperature profile by \cite{Lothringer+2018} and \cite{Arcangeli2021}.  

Our result for the mass-loss timescale $|M_p/\dot M_p|\approx 5\unit{Myr}$ is within a factor of 2 of that of \cite{Li+2010}, although this seems to be partly coincidence: their estimate for the density at the L1 point is more than two orders of magnitude less than ours, but they assume that the mass loss occurs through the entire Roche lobe, while we estimate the width of the ``nozzle'' around L1 following \cite{Jackson+2017}.  At the same time, our $\dot M_p$ is 50 times larger than the latter authors'.  A factor 10 is due to the differences in potential (our $n=1$ model vs. their point-mass approximation), the rest being due to differences in the temperature structure and to the assumed opacity at the transit radius. These things remain uncertain.  We also conclude that the mass-loss rate on the opposite side of the planet through the L2 point should be negligible, in contrast to the conclusions of \cite{Dwivedi+2019}, whose total mass-loss rate is $\sim 10^{-4}$ times ours because they allow for heating of the gas only by extreme ultraviolet radiation that carries only $\sim10^{-6}$ of the star's bolometric luminosity.


\section{Summary and Conclusion}

Using a version of the Ostriker-Mark method, we have computed polytropic models of ``planets" in Roche contact.
We are motivated by WASP 12b, because it is very close to this condition, is on a decaying orbit (so that if it is not yet in Roche contact, as we find, then it soon will be), and because it shows spectroscopic evidence for mass loss.
These methods, however, could be applied to other very-short-period hot Jupiters that have well-constrained densities, such as some of those in Table~2 of \cite{Jackson+2017}.

Because real jovian planets have atmospheres that are not on the same adiabat as their interiors, the notion of ``Roche contact'' is ambiguous.  
A more precise statement is that the equipotential and isobaric surfaces (these are not the same) of WASP 12b appear to be detached from the corresponding surfaces that pass through the inner saddle point of the effective potential ($L1$).  Our improved tidal models have allowed us to make more precise estimates of just how far WASP 12b is detached from its Roche lobe in this sense.  
We have also used these models to revisit the thermally-driven atmospheric mass-loss rate of the atmosphere, which we find to be comparable to the decay time of the orbit as measured by \cite{Yee+Winn+2020}---a few Myr.  
However, the planet will achieve Roche contact in only a few hundred thousand years.  
We have not attempted to analyze what may happen after that: the planet may disrupt, or the orbit may re-expand, as in some cataclysmic variables \citep{Warner2003}.

While an improvement over previous work as regards the tidal distortion and gravitational potential of the planet, our models remain highly idealized.  We have worked with polytropes rather than realistic equations of state and profiles of entropy and composition.  These remain quite uncertain, especially for highly ``inflated'' (low-density) cases such as WASP 12b.  The atmospheric structures that we have borrowed from previous work to estimate the mass-loss rate were computed without accounting for the tide; ideally one would treat the atmosphere and the tide self-consistently.  But it is not clear that the circulation of the atmosphere is well enough understood at present to justify the effort.

\appendix
\section{Estimate of the mass-loss rate}\label{sec:Mdot}
Our estimates are based on the semi-analytic approach of \cite{Jackson+2017}; in our notation, their eq.~(3) reads
\begin{equation}\label{eq:Mdot0}
\dot M = \rho_s e^{-\Psi_{\rm L1}/\cs^2-1/2} \times \cs\times 2\pi\cs^2\left(\frac{\partial^2\Psi}{\partial y^2}\frac{\partial^2\Psi}{\partial z^2}\right)^{-1/2}_{L1}
\end{equation}
Here $\rho_s$ is the mass density at the ``surface'' of the planet where $\Psi\equiv0$, $\Psi_{L1}$ is the value of the effective potential at the inner Lagrange point, and
$\cs=\sqrt{\kB T/\mu m_{\rm amu}}$ is the thermal speed for temperature $T$ and molecular weight $\mu$; $m_{\rm amu}$ is the atomic mass unit.  This formula is based on the following approximations:
\begin{enumerate}
    \item The gas temperature is presumed constant above the surface.
    \item The atmosphere is approximately hydrostatic below the sonic point, which lies near L1.
    \item Coriolis forces are neglected.
\end{enumerate}
The first two assumptions give the exponential in eq.~\eqref{eq:Mdot0}.  The last factor is the estimated cross section of the ``nozzle'' through which the flow passes, the second partial derivatives being evaluated at L1, where $\b{\nabla}\Psi=0$ and $\partial^2\Psi/\partial x^2<0$.  

Clearly, because of the exponential it is essential to have accurate models both for the effective potential and for the thermal state of the atmosphere.  \cite{Jackson+2017} evaluate the former in the usual restricted three-body approximation where the gravitational potential of the planet is that of a point mass.  We improve on this with our tidally-distorted polytropic models, and find that it makes a significant difference.  The indicated second partial derivatives in eq.~\eqref{eq:Mdot0}are quite similar to those derived by \cite{Jackson+2017} from the restricted 3-body problem, however.
For the temperature, \cite{Jackson+2017} use $T_{\rm eq}\equiv (F_*/2\sigma_{\textsc{sb}})^{1/4}$, where $F_*\equiv L_*/4\pi a^2$ is the insolation at the substellar point, as if the insolation were re-radiated uniformly over the ``day'' side without redistribution to the other hemisphere.  With the parameters of \cite{Collins+2017}, $F_*\approx 10^{10}\unit{erg\,cm^{-2}\,s^{-1}}$, so that $T_{\rm eq}\approx 3000\unit{K}$.
We improve on this using recent models for the temperature structure and circulation of the atmosphere by \cite{Lothringer+2018} and \cite{Arcangeli2021}.\footnote{We use \cite{Lothringer+2018}'s ``fiducial'' model for a generic hot Jupiter with insolation and surface gravity very similar to WASP 12b, as shown by their Figure~1.}  Since these atmospheres are not perfectly isothermal vertically, we replace the exponential
factor in eq.~\eqref{eq:Mdot0} (except for the $e^{-1/2}$, which represents the transition from hydrostatic to sonic flow) with eqs.~\eqref{eq:Hdef}-\eqref{eq:hydro1}, recast as
\begin{equation}\label{eq:VHE}
    \Psi(P_1)-\Psi(P_2) = \int\limits_{\ln P_1}^{\ln P_2} \frac{\kB T(P)}{\mu(P,T)m_p} \mathrm{d}\ln P\,
\end{equation}
with the understanding that the integral is to be taken vertically, perpendicular to the constant-pressure surfaces: since the entropy varies along the equipotentials, it is not possible for the atmosphere to be in exact equilibrium, but one assumes that hydrostatic equilibrium is better approximated vertically than horizontally because the atmospheric scale height $H\equiv\cs^2/g\ll R$.  The function $T(P)$ is taken from \cite{Lothringer+2018}'s fiducial model (their Figs.~2 \& 14), which is hotter than $T_{\rm eq}$---closer to $4000\unit{K}$---above the $10^{-4}\unit{bar}$ level, presumably because of the wavelength dependence of the opacity.  The molecular weight $\mu(P,T)$ varies with height because of dissociation of molecular hydrogen.  For this we assume LTE and the
partition function of \cite{Irwin1987} for $\mathrm{H}_2$.  The gas is predominantly molecular at the transit radius, but quickly becomes atomic with decreasing pressure.  We take $(X,Y,Z)=(0.71,0.27,0.02)$ for the abundances of hydrogen, helium, and metals, with $\bar\mu_Z\approx 16$.

Following \cite{Jackson+2017} and \cite{Howe+Burrows2012}, we identify the pressure level at the transit radius $R_T$ from the condition that the optical depth along the transit chord at the planet's limb should be $0.56$, i.e. $\tau_T\approx\kappa_\lambda\rho\sqrt{2\pi H R_T}=0.56$, where $\kappa_\lambda$, $\rho$, and $H\equiv\cs^2/g$ are evaluated at the tangent point.   We take $R_T=1.9\unit{R_J}$ and $M_p=1.47\unit{M_J}$ from \cite{Collins+2017}.  By scaling our $n=1$ model to these values, we find that the surface gravity at the tangent point is $g=0.913 GM_p/R_T^2=922\unit{cm/s^2}$ (whereas the gravity at the pole facing
the star is $410\unit{cm/s^2}$).  For the temperature at this point, we use $T=2500\unit{K}\approx (F_*/4\sigma_{\textsc{sb}})^{1/4}$, which is roughly consistent with the plots in \cite{Arcangeli2021}.  With these choices, hydrogen turns out to be overwhelmingly molecular at the tangent point, so $\mu\approx2.36$.
For the wavelength-dependent opacity $\kappa_\lambda$, we consider only the contribution of $\mathrm{H^-}$, calculating its volume mixing ratio from the formula provided by \cite{Lothringer+2018}, and its optical cross section from \cite{Ohmura+Ohmura1960}, evaluating this at $\lambda=0.62\unit{\mu m}$, the effective wavelength of the SDSS $r'$ band \citep{Collins+2017}.  We find $\kappa_{0.62\unit{\mu m}}(\mathrm{H^-})=0.002\unit{cm^2/g}$ and $P\approx 9\unit{mbar}$ at the tangent point.  This is quite important: from \cite{Lothringer+2018}, $\mathrm{H^-}$ appears to dominate the opacity at this wavelength, but a larger opacity would predict a smaller pressure and density at the tangent  point, and ultimately a smaller mass-loss rate, as discussed below.

The circulation models of \cite{Arcangeli2021} indicate that the temperature is almost constant with longitude at $P\gtrsim 1\unit{bar}$.  Therefore, by setting $P_2=1\unit{bar}$ in eq.~\eqref{eq:VHE}, we find the value of $P_1$ on the day side that corresponds to (i.e., lies on the same equipotential as) the pressure at the transit point by using the day-side temperature profile instead of the constant $2500\unit{K}$ that we assumed for the terminator; we also take into account the variation in molecular weight as described above.  This gives $P_{1,\rm day}=20\unit{mbar}$; presumably this would be slightly higher at the substellar point, where the temperatures will be higher than the average for the day side.  This equipotential is considered the ``surface'' of the planet in our tidal model, $\Psi(P_{\rm 1,day})=0$.  The potential at the Lagrange point $\tilde\Psi_{L1}=0.0988$ in Emden units, in which also the planet mass $\tilde M_p=42.1195$, transit radius $\tilde R_T=2.97396$, and squared orbital frequency $(1+\delta)\tilde\Omega^2=0.0076075$.
By scaling the latter three to the observed values from \cite{Collins+2017}, we find that the predicted potential difference between the planet surface and the Lagrange point is $\Psi_{L1}\approx(10.69\unit{km/s})^2$.  Using eq.~\eqref{eq:VHE} yet again, and including the factor $e^{-1/2}$ from eq.~\eqref{eq:Mdot0}, we have $P_{\rm L1}\approx 0.15\unit{mbar}$, $\rho_{\rm L1}\approx 5.7\times 10^{-10}\unit{g\,cm^{-3}}$, and $c_{\rm s,L1}\approx5.16\unit{km/s}$ (the hydrogen is fully atomic there).  In this way we obtain, finally, $\dot M\approx 1.7\times 10^{16}\unit{g\,s^{-1}}$,
corresponding to a mass-loss timescale $M_p/\dot M\approx 5\unit{Myr}$.  A fraction $(\dot M\Psi_{L1})(L_* R_p^2/a^2)\approx 0.003$ of the insolation is needed to drive this outflow.

To test the sensitivity of $\dot M$ to our assumptions about the internal structure of the planet, we have repeated the procedure described above for an $n=3$ rather than $n=1$ polytrope (scaled to the same mean density, etc.).  An $n=3$ model is of course much more centrally concentrated, and its potential is much better described in the restricted-three-body approximation. In this case, we find that
$\Psi_{\rm L1}\approx(13.28\unit{km\,s^{-1}})^2$, about $50\%$ larger than for the $n=1$ model.  As a result, the density at L1 is about one order of magnitude smaller than before, and so is the predicted mass-loss rate: $\dot M_{n=3}\approx 2\times10^{15}\unit{g\,s^{-1}}$.  

These two models (i.e. $n=1$ and $n=3$) probably bracket the truth as far as the dependence of $\dot M$ on the structure of the planet's interior is concerned.  However, $\dot M$ is also sensitive to the structure of the atmosphere.  In particular, the pressure at the transit point ($P_T$) is inversely proportional to the assumed opacity (including scattering as well as absorption).  \cite{Sing+2013} argue from transmission spectroscopy for the presence of $\mathrm{Al_2O_3}$ aerosols, and therefore $P_T$ between $0.02$ and $0.5$~mbar depending upon particle size. Also, \cite{Sing+2016} classify WASP-12b as ``hazy'' based on the difference between optical and infrared transit radii. If we put $P_T=0.02\unit{mbar}$ rather than $9\unit{mbar}$ as we found from $\mathrm{H^-}$ opacity alone, the day-side pressure on the same equipotential becomes $0.73\unit{mbar}$ rather than $20\unit{mbar}$, and $\dot M\to 1.2\times10^{15}\unit{g\,s^{-1}}$ (with our $n=1$ model) rather than $1.7\times 10^{16}\unit{g\,s^{-1}}$.  Notice that $\dot M$ varies sublinearly with $P_T$ because of the larger pressure scale height on the day side.  By measuring the secondary eclipse of WASP-12 in the optical with {\it HST}, however, \cite{Bell+2017} constrain the day-side geometric albedo of the planet to be less than $0.064$ at $97.5\%$ confidence, indicating very little scattering, and they reject the aerosol models of \cite{Sing+2013}.

We have used the same methods to estimate the mass-loss rate of the $n=1$ model through the L2 point, i.e. on the side of the planet opposite the star.  This is much smaller than $\dot M_{\rm L1}$.  The difference in potential between the Lagrange points and the ``surface" of the planet is $33\%$ larger for L2 than for L1.  If the night side had the same temperature structure as the day side, this would lower $\dot M_{\rm L2}$ by a factor of about 4.  But the temperature structure is not the same, of course.  While we could not find a detailed model of the night-side temperature profile, Fig.~12 of \cite{Arcangeli2021} offers color contour plots of the estimated temperature at five discrete pressure levels over the whole surface of the planet; the temperature at the anti-stellar point appears to decrease from $\approx 2500\unit{K}$ at 1~bar to $\approx 1100\unit{K}$ at 1~mbar.  If $T=1100\unit{K}$ at all $P\le 1\unit{mbar}$, then $\dot M_{\rm L2}\approx 10^2\unit{g\,s^{-1}}\sim 10^{-14}\dot M_{\rm L1}$.
This enormous contrast is due to the exponential sensitivity of eq.~\eqref{eq:Mdot0} to the thermal speed.  (In addition to the direct effect of the temperature on the thermal speed, there is an indirect effect via the molecular weight: at lower temperatures, the hydrogen remains molecular to much lower pressures, assuming LTE.)  At sufficiently low pressures, however, the cooling time of the gas will exceed the time that it takes to be advected from the day side, so we expect that the temperature should rise well above $1100\unit{K}$ at such low pressures.  Based on our own rough estimates (involving, e.g., cooling by rovibrational lines of CO), we are confident that the mass-loss rate through L2 should be at least two orders of magnitude smaller than through L1 insofar as eqs.~\eqref{eq:Mdot0}-\eqref{eq:VHE} apply.
\newpage
\bibliography{wasp12}
\bibliographystyle{aasjournal}
\end{document}